\documentclass[twocolumn]{revtex4}

\usepackage{amsmath}
\usepackage{amssymb}
\usepackage{graphicx}
\usepackage{epsfig}

\begin{document}

\title{On the equilibrium of self-gravitating neutrons, protons and electrons in $\beta$-equilibrium}

\author{Michael Rotondo}
\affiliation{Dipartimento di Fisica and ICRA, Sapienza Universit\`a di Roma, P.le Aldo Moro 5, I--00185 Rome, Italy}
\affiliation{ICRANet, P.zza della Repubblica 10, I--65122 Pescara, Italy}

\author{Jorge A. Rueda}
\affiliation{Dipartimento di Fisica and ICRA, Sapienza Universit\`a di Roma, P.le Aldo Moro 5, I--00185 Rome, Italy}
\affiliation{ICRANet, P.zza della Repubblica 10, I--65122 Pescara, Italy}

\author{Remo Ruffini}
\email{ruffini@icra.it}
\affiliation{Dipartimento di Fisica and ICRA, Sapienza Universit\`a di Roma, P.le Aldo Moro 5, I--00185 Rome, Italy}
\affiliation{ICRANet, P.zza della Repubblica 10, I--65122 Pescara, Italy}

\author{She-Sheng Xue}
\affiliation{Dipartimento di Fisica and ICRA, Sapienza Universit\`a di Roma, P.le Aldo Moro 5, I--00185 Rome, Italy}
\affiliation{ICRANet, P.zza della Repubblica 10, I--65122 Pescara, Italy}

\date{\today}

\begin{abstract}
We have recently proved the impossibility of imposing the condition of local charge neutrality in a  self-gravitating system of degenerate neutrons, protons and electrons in $\beta$-equilibrium. The coupled system of the general relativistic Thomas-Fermi equations and the Einstein-Maxwell equations have been shown to supersede the traditional Tolman-Oppenheimer-Volkoff equations. Here we present the Newtonian limit of the new equilibrium equations. We also extend the treatment to the case of finite temperatures and finally we give the explicit demonstration of the constancy of the Klein potentials in the case of finite temperatures generalizing the condition of constancy of the general relativistic Fermi energies in the case of zero temperatures.
\end{abstract}

\maketitle

\section{Introduction}\label{sec:1}

It is well known that the classic work of Oppenheimer and Volkoff \cite{oppenheimer39} addresses the problem of neutron star equilibrium configurations composed only of neutrons. For the more general case when protons and electrons are also considered, in nearly all of the scientific literature on neutron stars it is assumed that the condition of local charge neutrality applies identically to all points of the equilibrium configuration (see e.g.~\cite{nsbook}). Consequently, the corresponding solutions of the Einstein equations for a non-rotating neutron star, following the work of Tolman \cite{tolman39} and of Oppenheimer and Volkoff \cite{oppenheimer39}, have been systematically applied. However, the necessity of processes leading to electrodynamical phenomena during the gravitational collapse to a black hole \cite{physrep} suggests a critical reexamination of the current treatment of neutron stars. In this regard in a set of recent articles (see e.g.~\cite{PRC2011,PLB2011}) we have developed the first steps toward a new consistent treatment for the description of neutron stars overcoming the traditional Tolman-Oppenheimer-Volkoff equations.

We have recently proved the impossibility of imposing the condition of local charge neutrality in a self-gravitating system of degenerate neutrons, protons and electrons in $\beta$-equilibrium \cite{PLB2011}. It has been there discussed that the traditional approach for the description of neutron stars adopting the condition of local charge neutrality is inconsistent with the Einstein-Maxwell equations and with micro-physical conditions of equilibrium within the quantum statistics. We emphasize the basic role of the constancy of the general relativistic Fermi energy of each species. Consequently, the traditional Tolman-Oppenheimer-Volkoff system of equations has been superseded by the coupled system of the general relativistic Thomas-Fermi equations and the Einstein-Maxwell equations. We have been guided in this new approach by the  generalization of the Feynman-Metropolis-Teller treatment of compressed atoms to the relativistic regimes \cite{PRC2011}. There, using the existence of scaling laws \cite{ruffini2008,popov2010,popov2011}, the results were extended from heavy nuclei to the case of globally neutral nuclear matter cores of stellar dimensions with mass numbers $A\approx (m_{Planck}/m_n)^3\approx10^{57}$ or $M_{core}\approx M_{\odot}$.

In the present article we first recall in Sec.~\ref{sec:2} the new system of general relativistic  Einstein-Maxwell-Thomas-Fermi governing a self-gravitating system of neutrons, protons and electrons in $\beta$-equilibrium. We then address three new aspects of the problem: 1) we study the Newtonian limit of the equilibrium equations (see Sec. III); 2) we consider the proper generalization of the above treatment to the case of finite temperatures (see Sec. IV); 3) we generalize and prove the constancy of the Klein potentials of each species generalizing the condition of constancy of the general relativistic Fermi energies encountered in the case of zero temperatures (see Sec.~\ref{sec:4}). We finally show that the thermal effects do not affect the electrodynamical structure of the equilibrium configurations and the results are qualitatively and quantitatively quite similar to the ones obtained with the degenerate approximation (see Sec.~\ref{sec:4}). 

\section{ Einstein-Maxwell-Thomas-Fermi equations in the degenerate case }\label{sec:2}

Following \cite{PLB2011} we consider the equilibrium configurations of a degenerate gas of neutrons, protons and electrons with total matter energy density and total matter pressure
\begin{eqnarray}
{\cal E} &=& \sum_{i=n,p,e} \frac{2}{(2 \pi \hbar)^3} \int_0^{P^F_i} \epsilon_i(p)\,4 \pi p^2 dp\, ,\label{eq:eos1}\\
P &=& \sum_{i=n,p,e} \frac{1}{3} \frac{2}{(2 \pi \hbar)^3} \int_0^{P^F_i} \frac{p^2}{\epsilon_i(p)}\,4 \pi p^2 dp \, ,\label{eq:eos2}
\end{eqnarray}
where $\epsilon_i(p) = \sqrt{c^2 p^2+m^2_i c^4}$ is the relativistic single particle energy and $P^F_i$ denote the Fermi momentum, related to the particle number density $n_i$ by $n_i = (P^F_i)^3/(3 \pi^2 \hbar^3)$.

Introducing the metric for a spherically symmetric non-rotating configuration
\begin{equation}\label{eq:metric}
ds^2 = {\rm e}^{\nu(r)} c^2 dt^2 - {\rm e}^{\lambda(r)} dr^2 - r^2 d\theta^2 - r^2 \sin^2 \theta d\varphi^2\, ,
\end{equation}
the extension to general relativity of the Thomas-Fermi equilibrium condition on the generalized Fermi energies 
\begin{equation}\label{eq:electroneq}
E^F_i = {\rm e}^{\nu/2} \mu_i - m_i c^2 +q_i V \, ,
\end{equation}
($i=n,p,e$, $q_i=0,e,-e$) and the condition of $\beta$-equilibrium between neutrons, protons and electrons
\begin{equation}\label{eq:betaeq}
E^F_n + m_n c^2 = E^F_p + m_p c^2+ E^F_e + m_e c^2 \, ,
\end{equation}
the full system of equations composed by the Einstein-Maxwell-Thomas-Fermi equations can be written as (see \cite{PLB2011} for details)
\begin{eqnarray}
&&M' = 4 \pi r^2 \frac{{\cal E}}{c^2} - \frac{4 \pi r^3}{c^2} {\rm e}^{-\nu/2} \hat{V}' \Bigg\{ n_p \nonumber \\
&&- \frac{{\rm e}^{-3 \nu/2}}{3 \pi^2}[\hat{V}^2 + 2 m_e c^2 \hat{V} - m^2_e c^4 ({\rm e}^{\nu}-1)]^{3/2}\Bigg\}\, ,\label{eq:Gab12}\\
&&\nu' = \frac{2 G}{c^2} \frac{4 \pi r^3 P/c^2 + M - r^3 E^2/c^2}{r^2 \left(1-\frac{2 G M}{c^2 r} + \frac{G r^2}{c^4} E^2 \right)}
,\label{eq:Gab22}\\
&&E^F_e={\rm e}^{\nu/2}\mu_e-m_e c^2 -eV = {\rm constant},\label{eq:efe2}\\
&&E^F_p ={\rm e}^{\nu/2}\mu_{p}-m_p c^2 + eV = {\rm constant}, \label{eq:efp2}\\
&&E^F_n = E^F_e + E^F_p - (m_n-m_e-m_p) c^2, \label{eq:efn2}\\
&&\hat{V}'' + \frac{2}{r}\hat{V}' \left[ 1 - \frac{r (\nu'+\lambda')}{4}\right] = - 4 \pi \alpha \hbar c \, {\rm e}^{\nu/2} {\rm e}^{\lambda} \Bigg\{ n_p \nonumber \\
&&- \frac{{\rm e}^{-3 \nu/2}}{3 \pi^2}[\hat{V}^2 + 2 m_e c^2 \hat{V} - m^2_e c^4 ({\rm e}^{\nu}-1)]^{3/2}\Bigg\}\, .\label{eq:GRTF2}
\end{eqnarray}
where $e$ is the fundamental charge, $\alpha$ is the fine structure constant, $V$ is the Coulomb potential, $\mu_i = \partial {\cal E}/\partial n_i = \sqrt{c^2 (P^F_i)^2+m^2_i c^4}$ is the free-chemical potential of particle-species, $\lambda(r)$ is the metric function related to the mass $M(r)$ and the electric field $E(r) = -{\rm e}^{-(\nu+\lambda)/2} V'$ (a prime stands for radial derivative) through 
\begin{equation}\label{eq:lambda}
{\rm e}^{-\lambda} = 1 - \frac{2 G M(r)}{c^2 r} + \frac{G}{c^4} r^2 E^2(r)\, .
\end{equation}
and $\hat V=E_e^F+eV$.

\begin{figure}
\centering
\includegraphics[width=\columnwidth,clip]{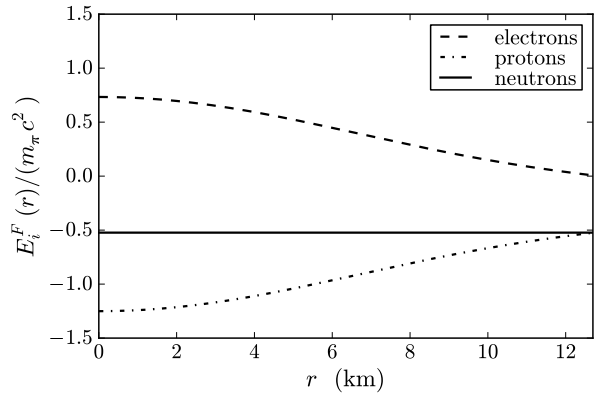}
\caption{Fermi energies for neutrons, protons and electrons in units of the pion rest-mass energy for a locally neutral configuration with central density $\rho(0) = 3.94\rho_{\rm nuc}$, where $\rho_{\rm nuc} = 2.7\times 10^{14}$ g cm$^{-3}$ denotes the nuclear density.}\label{fig:1}
\end{figure}

\begin{figure*}
\centering
\begin{tabular}{lr}
\includegraphics[width=\columnwidth,clip]{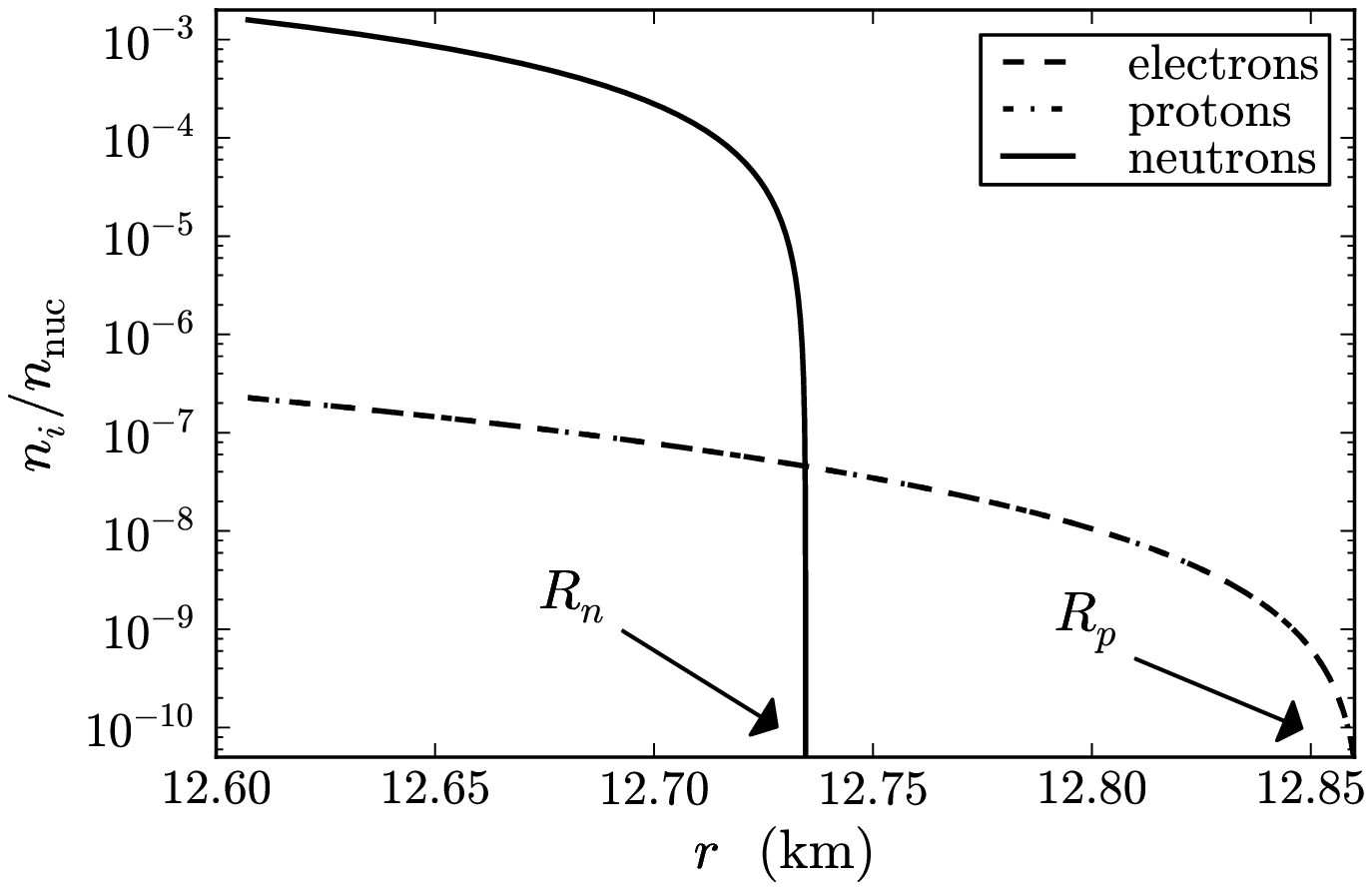} & \includegraphics[width=\columnwidth,clip]{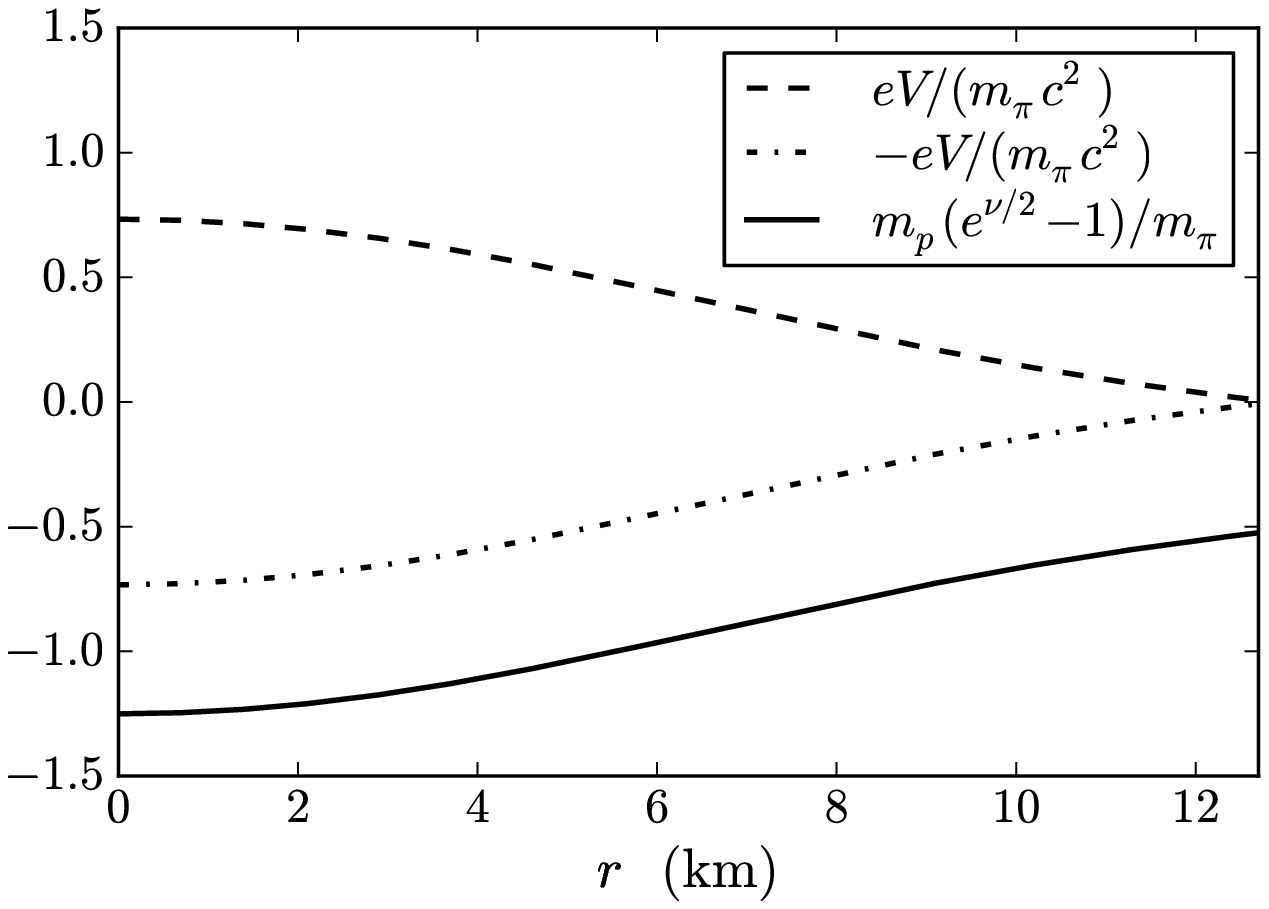}
\end{tabular}
\caption{Left panel: particle number density approaching the boundary of the configuration in units of the nuclear density $n_{\rm nuc} \simeq 1.6\times 10^{38}$ cm$^{-3}$. Right panel: proton and electron Coulomb potential in units of the pion rest-mass energy $eV/(m_\pi c^2)$ and $-eV/(m_\pi c^2)$ respectively and the proton gravitational potential in units of the pion mass $m_p ({\rm e}^{\nu/2}-1)/m_\pi$.}\label{fig:fig2}
\end{figure*}

The condition $n_e=n_p$ often adopted in literature is not consistent with Eqs.~(\ref{eq:efe2}) and (\ref{eq:efp2}) (see Fig.~\ref{fig:1}) therefore we consider the equilibrium configurations fulfilling only global charge neutrality (details are given in \cite{PLB2011}). We solve self-consistently Eq.~(\ref{eq:Gab12}) and (\ref{eq:Gab22}) for the metric, Eqs.~(\ref{eq:efe2})--(\ref{eq:efn2}) for the equilibrium of the three degenerate fermion species and for the $\beta$-equilibrium. The crucial equation relating the proton and the electron distributions is then given by the general relativistic Thomas-Fermi equation (\ref{eq:GRTF2}).
The boundary conditions are: for Eq.~(\ref{eq:Gab12}) the regularity at the origin: $M(0)=0$, for Eqs.~(\ref{eq:efe2})--(\ref{eq:efn2}) a given value of the central density, and for Eq.~(\ref{eq:GRTF2}) the regularity at the origin $n_e(0)=n_p(0)$, and a second condition at infinity which results in an eigenvalue problem determined by imposing the global charge neutrality conditions
\begin{equation}\label{eq:bound1}
\hat V (R_e) = E^F_e\, ,\qquad \hat V'(R_e) = 0\, ,
\end{equation}
at the radius $R_e$ of the electron distribution defined by 
\begin{equation}\label{eq:bound2}
P^F_e (R_e) = 0\, ,
\end{equation}
from which follows
\begin{eqnarray}\label{eq:bound3}
E^F_e &=& m_e c^2 {\rm e}^{\nu(R_e)/2} - m_e c^2 \nonumber \\ 
&=& m_e c^2 \sqrt{1-\frac{2 G M(R_e)}{c^2 R_e}} - m_e c^2\, .
\end{eqnarray}
Then the eigenvalue problem consists in determining the gravitational potential and the Coulomb potential at the center of the configuration that satisfy the conditions (\ref{eq:bound1})--(\ref{eq:bound3}) at the boundary.
The solution for the density, the gravitational potential and electric potential are shown in Fig.~(\ref{fig:fig2}) for a configuration with central density $\rho(0)=3.94\rho_{\rm nuc}$.
One particular interesting new feature is the approach to the boundary of the configuration: three different radii are present corresponding to distinct radii at which the individual particle Fermi pressures vanish. The radius $R_e$ for the electron component corresponding to $P^F_e (R_e) = 0$, the radius $R_p$ for the proton component corresponding to $P^F_p (R_p) = 0$ and the radius $R_n$ for the neutron component corresponding to $P^F_n (R_n) = 0$. 
For the configuration of Fig.~\ref{fig:fig2} we found $R_n \simeq 12.735$ km, $R_p \simeq 12.863$ km and $R_e \simeq R_p + 10^3 \lambda_e$ where $\lambda_e=\hbar/(m_e c)$ denotes the electron Compton wavelength. We find that the electron component follows closely the proton component up to the radius $R_p$ and neutralizes the configuration at $R_e$ without having a net charge, contrary to the results e.g in \cite{olson78}. 

\begin{figure*}
\centering
\begin{tabular}{lr}
\includegraphics[width=\columnwidth,clip]{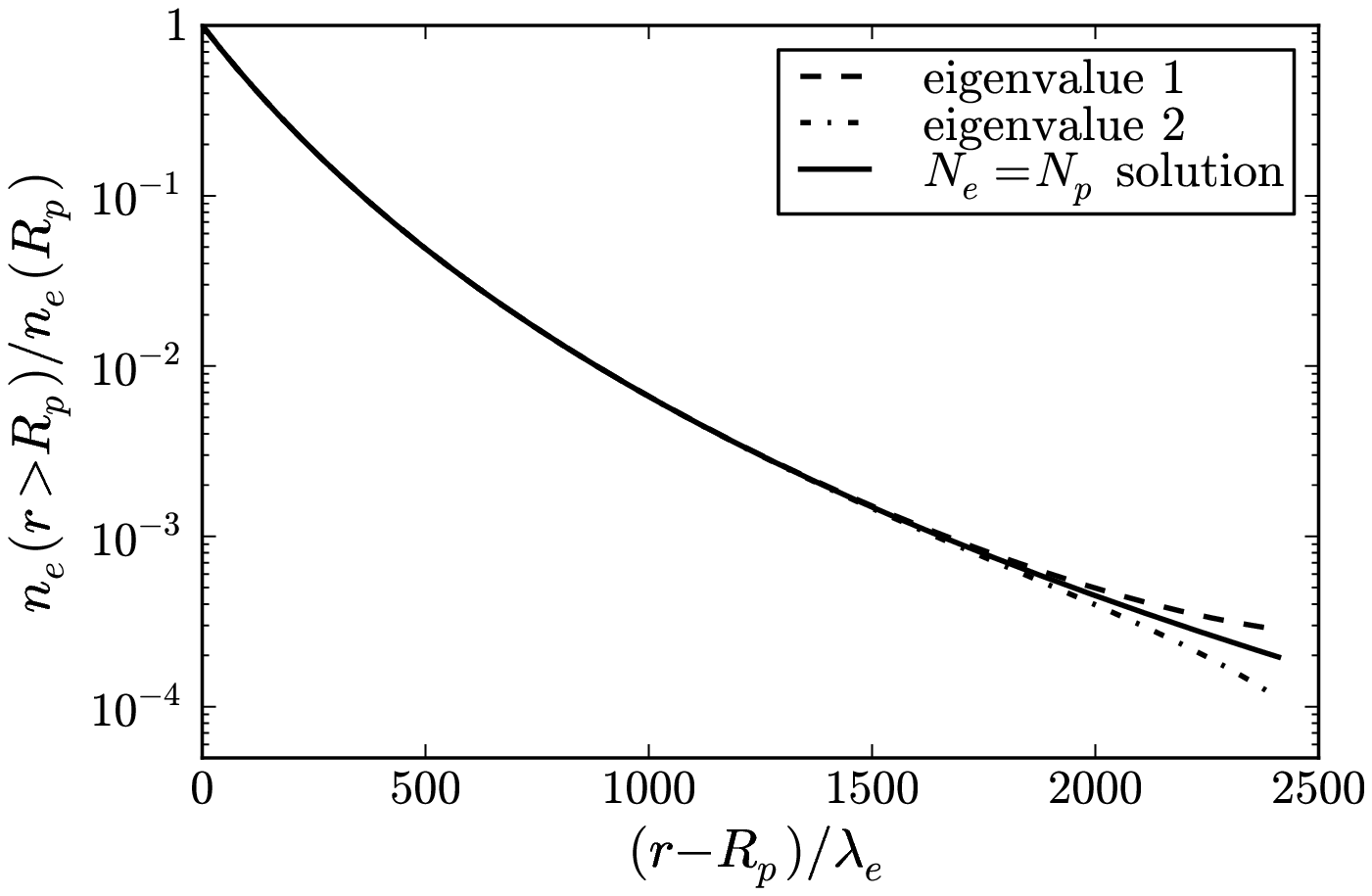} & \includegraphics[width=\columnwidth,clip]{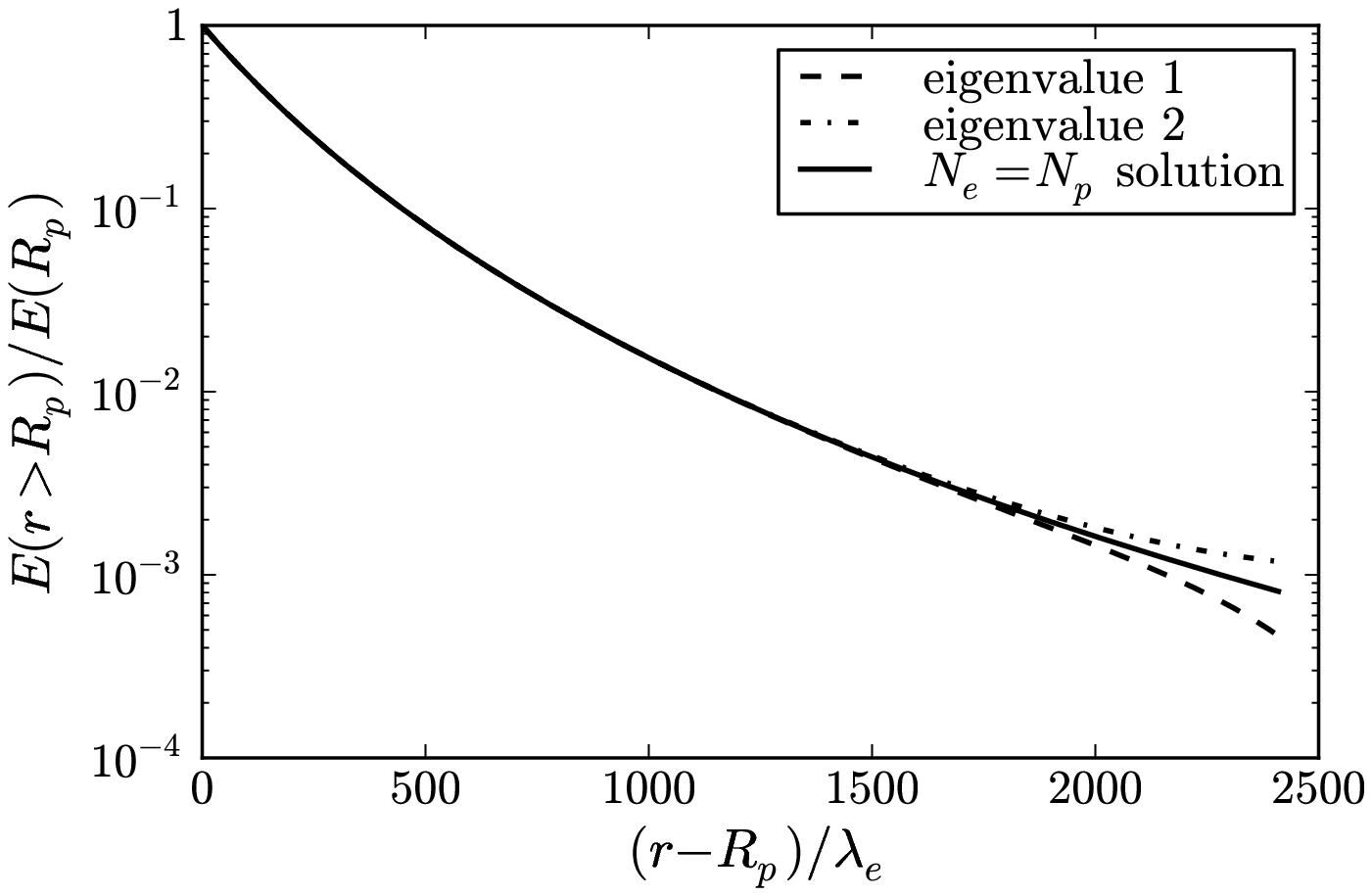}
\end{tabular}
\caption{Left panel: electron number density for $r\geq R_p$ normalized to its value at $r=R_p$. Right panel: electric field for $r\geq R_p$ normalized to its value at $r=R_p$. We have also shown the behavior of the solution for two different eigenvalues of the general relativistic Thomas-Fermi equation (\ref{eq:GRTF2}) close to the one which gives the globally neutral configuration (see \cite{PLB2011} for details).}\label{fig:fig3}
\end{figure*}

\section{Newtonian limit}\label{sec:3}

Despite the fact that the strong gravitational field of neutron stars requires a general relativistic treatment, it is interesting to explore the Newtonian limit of all the above considerations. This can help to elucidate if the gravito-electromagnetic effects we have found are of general relativistic nature or to prove their validity in a Newtonian regime.

The Newtonian limit of the equilibrium equations can be obtained by the weak-field non-relativistic limit. We expand the gravitational potential at first-order $e^{\nu/2} \approx 1 + \Phi$, where the Newtonian gravitational potential is $\Phi (r) = \nu(r)/2$. In the non-relativistic mechanics limit $c\to \infty$, the particle chemical potential becomes $\mu_i \to \tilde{\mu}_i + m_i c^2$, where $\tilde{\mu}_i = (P^F_i)^2/(2 m_i)$ denotes the non-relativistic free-chemical potential. Applying these considerations, the electron equilibrium law (\ref{eq:electroneq}) becomes
\begin{equation}\label{eq:efeNewt}
E^{F,{\rm Newt}}_e= \tilde{\mu}_e + m_e \Phi -e V \, ,
\end{equation}
which is the classical condition of thermodynamic equilibrium in presence of external gravitational an electrostatic fields (see e.g.~\cite{klein49} and \cite{landaubook}) applied to a gas of electrons.

In the weak-field non-relativistic limit, the Einstein-Maxwell equations (\ref{eq:Gab12})--(\ref{eq:GRTF2}) become
\begin{align}
&M' = 4 \pi r^2 \rho(r)\, ,\label{eq:Gablimit1}\\
&\Phi' = \frac{G M}{r^2}\, ,\label{eq:Gablimit2}\\
&P' = - \frac{G M}{r^2} \rho - \Bigg[n_p -\frac{(2 m_e)^{3/2}}{3 \pi^2 \hbar^3}(\hat{V} - m_e \Phi)^{3/2}\Bigg] \hat{V}'\, ,\label{eq:Gablimit3} \\
&\hat{V}'' + \frac{2}{r}\hat{V}' = - 4 \pi e^2 \,\Bigg[ n_p - \frac{(2 m_e)^{3/2}}{3 \pi^2 \hbar^3}(\hat{V} - m_e \Phi)^{3/2}\Bigg]\, ,\label{eq:Gablimit4}
\end{align}
where $\rho$ in this case is the rest-mass density
\begin{equation}
\rho = \sum_{i=n,p,e} m_i n_i\, .
\end{equation}

\begin{figure}
\centering
\includegraphics[width=\columnwidth,clip]{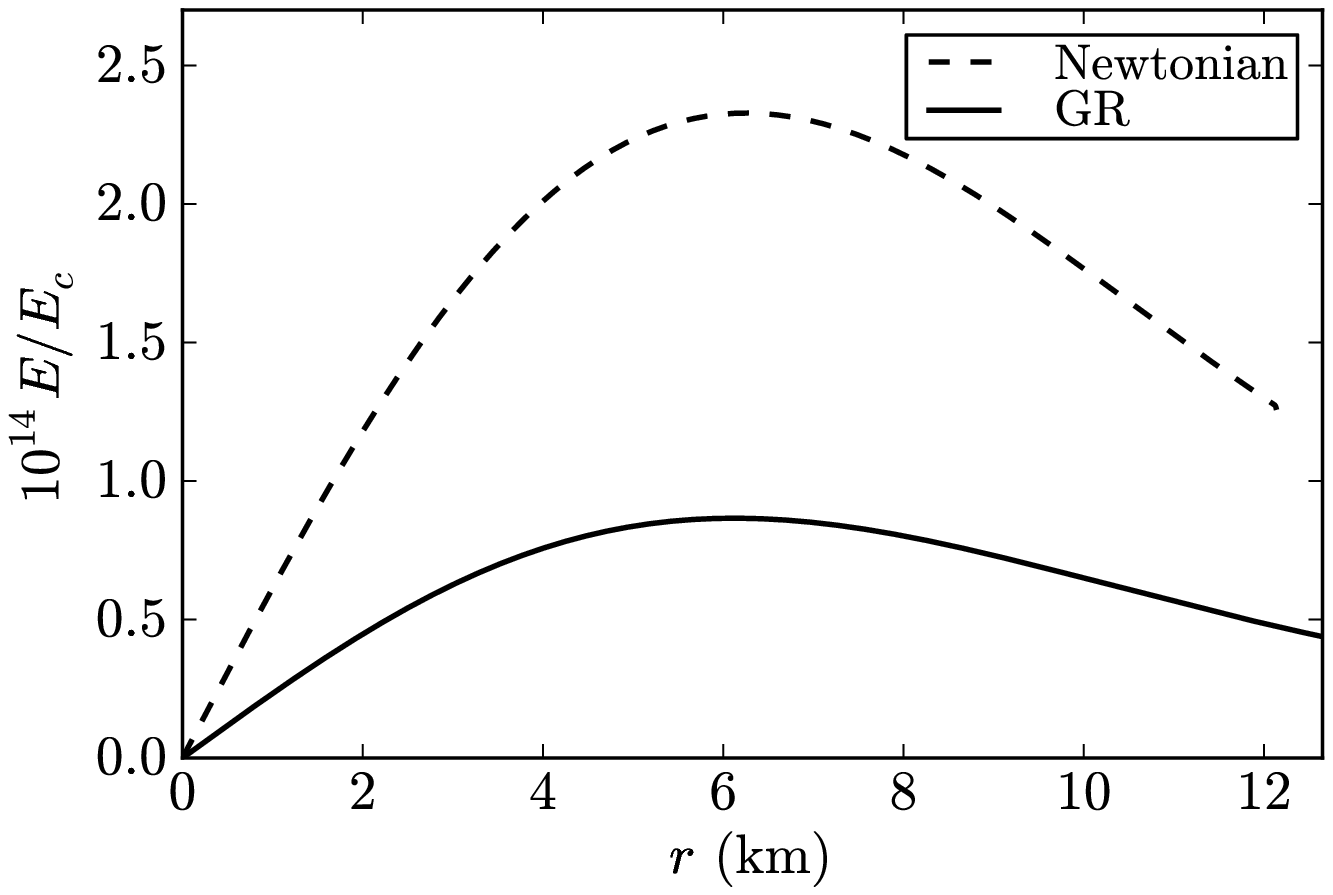}
\caption{Electric field (multiplied by $10^{14}$) in units of the critical field $E_c=m^2_e c^3/(e \hbar) \approx 10^{16}$ Volt/cm in the region $r<R_n$ both for the Newtonian and the General Relativistic configurations. The central density of both systems is $\rho(0)=3.94\rho_{\rm nuc}$ where $\rho_{\rm nuc}=2.7\times 10^{14}$ g cm$^{-3}$ is the nuclear density.}\label{fig:EGRNewtonian}
\end{figure}

The solution of Eqs.~(\ref{eq:efeNewt}), (\ref{eq:Gablimit1})--(\ref{eq:Gablimit4}) together with the $\beta$-equilibrium condition
\begin{equation}
E^{F,{\rm Newt}}_n = E^{F,{\rm Newt}}_p + E^{F,{\rm Newt}}_e\, ,
\end{equation}
leads to qualitatively similar electrodynamical properties as the one obtained in the general relativistic case. In Fig.~\ref{fig:EGRNewtonian} we show the electric field in the region $r<R_n$ ($R^{\rm Newt}_n < R^{\rm GR}_n$) both for the Newtonian as well as for the General Relativistic configuration for the given central density $\rho(0)=3.94\rho_{\rm nuc}$. From the quantitative point of view, the electric field of the Newtonian configuration is larger than the electric field of the general relativistic configuration.

\section{Finite temperature effects}\label{sec:4}

The above results have been obtained within the zero temperature approximation. It is worth to recall that temperatures of the order of $\sim 10^6$ K expected to exist at the surface of old neutron stars \cite{crab1,crab2}, or temperatures of $10^7-10^8$ K which could, in principle, exist in neutron star interiors, should not affect the considerations here introduced. For neutron stars, the Fermi temperature 
\begin{equation}\label{eq:Tfermi}
T^F_i = \frac{\mu_i-m_i c^2}{k}\, ,
\end{equation}
where $k$ is the Boltzmann constant, can be as large as $\sim 10^{12}$ K for electrons, $\sim 10^{11}$ K for protons and $\sim 10^{13}$ K for neutrons for typical central densities of neutron stars, which means that neutron star interiors are, at a high degree of accuracy, degenerate systems. However, the thermal energy associated to temperatures $T \sim 10^8$ K, $E_{\rm th} \sim (4\pi/3) R^3 a T^4 \sim 10^{38}$ erg, is much larger than the Coulomb energy $E_{\rm C} \sim (1/6) R^3 E^2 \sim 10^{16}$ erg, given by the internal electric field here considered (see Fig.~\ref{fig:EGRNewtonian}). Here $R$ is the radius of the configuration and $a=\sigma/c$ being $\sigma$ the Stefan-Boltzmann constant. It can be then of interest to ask the question if our electrodynamical structure will still occur in presence of thermal effects.

In this more general case, the equation of state given by Eqs.~(\ref{eq:eos1}) and (\ref{eq:eos2}), must be abandoned and replaced by
\begin{eqnarray}
{\cal E} &=& \sum_{i=n,p,e} \frac{2}{(2 \pi \hbar)^3} \int_0^\infty \tilde{\epsilon}_i(p) f_i (p)\, 4 \pi p^2 dp\, ,\label{eq:eos1T}\\
P &=&\sum_{i=n,p,e} \frac{1}{3} \frac{2}{(2 \pi \hbar)^3} \int_0^\infty \frac{p^2 f_i (p)}{\tilde{\epsilon}_i(p) + m_i c^2} \, 4 \pi p^2 dp\, ,\label{eq:eos2T}
\end{eqnarray}
where 
\begin{equation}
f_i (p) =  \frac{1}{1+e^{\frac{\tilde{\epsilon}_i(p)-\tilde{\mu}_i}{k T}}}\, ,
\end{equation}
is the Fermi-Dirac fermion distribution function which gives the particle number density $n_i$ 
\begin{eqnarray}
n_i= \frac{2}{(2 \pi \hbar)^3} \int_0^\infty  f_i (p)\, 4 \pi p^2  dp,\label{eq:eos2Ta}\, ,
\end{eqnarray}
where, for numerical purposes, the free single particle energy $\tilde{\epsilon}_i(p) = \epsilon_i(p) - m_i c^2 = \sqrt{c^2 p^2+m^2_i c^4}-m_i c^2$ and the free particle chemical potential $\tilde{\mu_i}$, have been defined subtracting the particle rest mass-energy.

We turn now to demonstrate the constancy of the Klein potentials throughout the configuration. The equation of state (\ref{eq:eos1T})--(\ref{eq:eos2T}) satisfies the thermodynamic law
\begin{equation}\label{eq:thermolaw}
\mathcal{E}+P-T s =\sum_{i=n,p,e}n_i \mu_i\, ,
\end{equation}
where $s = S/V$ is the entropy per unit volume and $\mu_i= \partial\mathcal{E}/\partial n_i$ is the free-chemical potential of the $i$-specie. At zero-temperature $T=0$, $\mu_i =\sqrt{(P_i^F)^2+\tilde{m}^2_i}$ and $n_i=(P_i^F)^3/(3 \pi^2)$, where $P_i^F$ denotes the Fermi momentum of the $i$-specie.

From Eq.~(\ref{eq:thermolaw}) follows the Gibbs-Duhem relation
\begin{equation}\label{Eq:GGDR}
dP=\sum_{i=n,p,e} n_i d\mu_i+ s dT\, ,
\end{equation}
which can be rewritten as
\begin{equation}\label{Eq:GGDR2}
dP=\sum_{i=n,p,e} n_i d\mu_i  + \left({\cal E}+P-\sum_{i=n,p,e} n_i \mu_i \right) \frac{dT}{T}\, .
\end{equation}

Introducing the metric given by Eq.~(\ref{eq:metric}) the Einstein-Maxwell system of equations is
\begin{eqnarray}
&&M' = 4 \pi r^2 \frac{{\cal E}}{c^2} - \frac{4 \pi r^3}{c^2} {\rm e}^{-\nu/2} V' ( n_p-n_e)\, ,\label{eq:Gab1a}\\
&&\nu' = \frac{2 G}{c^2} \frac{4 \pi r^3 P/c^2 + M - r^3 E^2/c^2}{r^2 \left(1-\frac{2 G M}{c^2 r} + \frac{G r^2}{c^4} E^2 \right)}
,\label{eq:Gab2a}\\
&&P'+\frac{\nu'}{2} ({\cal E} + P) = - {\rm e}^{-\nu/2} V' (n_p-n_e)\, ,\label{eq:TOVa}\\
&&V'' + \frac{2}{r}V' \left[ 1 - \frac{r (\nu'+\lambda')}{4}\right] = - 4 \pi \alpha \hbar c \, {\rm e}^{\nu/2} {\rm e}^{\lambda} ( n_p\nonumber \\
&&-n_e)\, .\label{eq:GRTFa}
\end{eqnarray} 

Using the Gibbs-Duhem relation (\ref{Eq:GGDR2}) the energy-momentum conservation equation (\ref{eq:TOVa}) can be rewritten as
\begin{eqnarray}
&&{\rm e}^{\nu/2}\sum_{i=n,p,e} n_i \left( d\mu_i - \frac{dT}{T} \mu_i\right) + ({\cal E}+ P){\rm e}^{\nu/2}\left(\frac{dT}{T}\right. \nonumber \\ && \left.+ \frac{1}{2} d\nu \right) + e (n_p-n_e) dV=0\, . \label{eq:TOV1a}
\end{eqnarray}

The Tolman isothermal condition \cite{tolman30} (see also \cite{klein49}) demands the constancy of the gravitationally red-shifted temperature
\begin{equation}\label{eq:Tcons}
\frac{dT}{T} + \frac{1}{2} d\nu = 0\, , \qquad {\rm or}\qquad T_{\infty}={\rm e}^{\nu/2} T = {\rm constant}\, ,
\end{equation}
which can be used into Eq.~(\ref{eq:TOV1a}) to obtain
\begin{equation}\label{Eq:TOVPINTRO}
\sum_{i=n,p,e} n_i d({\rm e}^{\nu/2} \mu_i) + e (n_p-n_e) dV=0\, .
\end{equation}

We now introduce the generalized chemical potentials, or Klein potentials, for electrons $E_e$, protons $E_p$ and neutrons $E_n$
\begin{eqnarray}
\label{Eq:gec1} E_e&=&{\rm e}^{\nu/2}\mu_e -m_ec^2-eV\, ,\\
\label{Eq:gec2}E_p &=& {\rm e}^{\nu/2}\mu_p-m_pc^2+eV\, ,\\
\label{Eq:gec12}E_n&=&{\rm e}^{\nu/2}\mu_n-m_nc^2\, ,
\end{eqnarray}
which in the zero temperature limit are the  generalized Fermi energies for electrons $E_e=E_e^F$, neutrons $E_n=E_n^F$ and protons $E_p=E_p^F$ introduced in Sec. II (see Eq.~(\ref{eq:electroneq})). Using Eqs.~(\ref{Eq:gec1}), (\ref{Eq:gec2}) and (\ref{Eq:gec12}), Eq.~(\ref{Eq:TOVPINTRO}) becomes
\begin{equation}\label{Eq:2pas}
\sum_{i=n,p,e} n_i dE_i=0\, ,
\end{equation}
which leads for independent and non-zero particle number densities $n_i \neq 0$ to the constancy of the Klein potentials (\ref{Eq:gec1})--(\ref{Eq:gec12}) for each particle-species, i.e.
\begin{eqnarray}
E_e &=& {\rm e}^{\nu/2}\mu_e -m_ec^2-eV={\rm constant}\, ,\label{eq:ef1}\\
E_{p} &=& {\rm e}^{\nu/2}\mu_{p} -m_pc^2+eV = {\rm constant}\, ,\label{eq:ef2} \\
E_{n} &=& {\rm e}^{\nu/2}\mu_{n} -m_nc^2 = {\rm constant}\, .\label{eq:ef3}
\end{eqnarray}

In the zero temperature limit the constancy of the Klein potential of each particle-specie becomes the constancy of the generalized Fermi energies  introduced in Sec.~\ref{sec:2} (see Eqs.~(\ref{eq:efe2}) and (\ref{eq:efp2})). 
This is a crucial point because, as discussed in \cite{PLB2011}, the constancy of the generalized Fermi energies proves the impossibility of having a self-consistent configuration fulfilling the condition of local charge neutrality and $\beta$-equilibrium (see e.g.~Fig.~\ref{fig:1}). Further, as shown in \cite{PLB3011}, the constancy of the Klein potentials holds in the more general case when the strong interactions between nucleons are taken into account. 

Therefore, introducing the new dimensionless variables $\eta_i = \tilde{\mu}_i/(k T)$ and $\beta_i = k T/(m_ic^2)$,  the new set of Einstein-Maxwell-Thomas-Fermi equations generalizing the system (\ref{eq:Gab12})--(\ref{eq:GRTF2}) to the case of finite temperatures is
\begin{eqnarray}
&&M' = 4 \pi r^2 \frac{{\cal E}}{c^2} - \frac{4 \pi r^3}{c^2} {\rm e}^{-\nu/2} \hat{V}' (n_p -n_e)\, ,\label{eq:Gab1T}\\
&&\nu' = \frac{2 G}{c^2} \frac{4 \pi r^3 P/c^2 + M - r^3 E^2/c^2}{r^2 \left(1-\frac{2 G M}{c^2 r} + \frac{G r^2}{c^4} E^2 \right)}
,\label{eq:Gab2T}\\
&&E_e=m_e c^2 {\rm e}^{\nu/2} (1 + \beta_e \eta_e) - m_e c^2 - e V \nonumber \\
&&= {\rm constant},\label{eq:efeT}\\
&&E_p =m_p c^2 {\rm e}^{\nu/2} (1 + \beta_p \eta_p) - m_p c^2 + e V \nonumber \\
&&= {\rm constant}, \label{eq:efpT}\\
&&E_n = E_e + E_p - (m_n-m_e-m_p) c^2, \label{eq:efnT}\\
&&\hat{V}'' + \frac{2}{r}\hat{V}' \left[ 1 - \frac{r (\nu'+\lambda')}{4}\right] = - 4 \pi \alpha \hbar c \, {\rm e}^{\nu/2} {\rm e}^{\lambda} (n_p \nonumber \\
&&-n_e)\, ,\label{eq:GRTFT}\\
&&{\rm e}^{\nu/2} \beta_i = {\rm constant}\, ,\qquad i = n, p, e\, . \label{eq:betaseq}
\end{eqnarray}
where Eq.~(\ref{eq:efnT}) is the condition of $\beta$-equilibrium between neutrons, protons and electrons, and the number density of the $i$-specie is given by
\begin{equation}
n_i = \frac{2^{1/2}m^3_i c^3}{\pi^2\hbar^3}\beta^{3/2}_i (F^i_{1/2}+\beta_i F^i_{3/2})\, ,
\end{equation}
where we have introduced the relativistic Fermi-Dirac integrals of order $j$ 
\begin{equation}
F^i_j = F_j(\eta_i,\beta_i) = \int_0^\infty \frac{x^j \left(1 + \frac{1}{2} \beta_i x\right)^{1/2}}{1 + e^{x-\eta_i}} dx\, .
\end{equation}

\begin{figure}
\centering
\includegraphics[width=\columnwidth,clip]{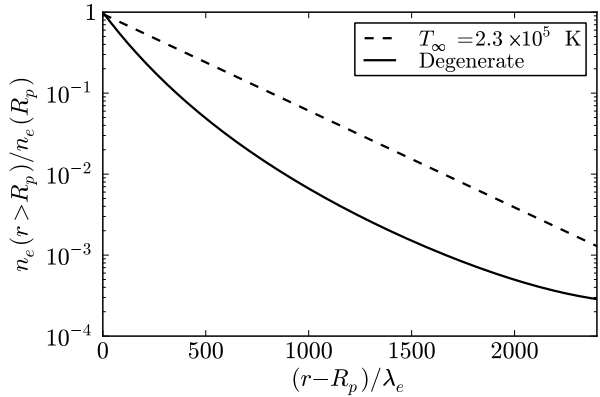}
\caption{Electron number density for $r\geq R_p$ normalized to its value at $r=R_p$ both for $T=0$ K (degenerate case) and for a finite temperature of $T_\infty=2.3\times 10^5$ K.}\label{fig:neTdeg}
\end{figure}

We have integrated numerically the system of equations (\ref{eq:Gab1T})--(\ref{eq:betaseq}) for given temperatures $T_\infty \neq 0$. As expected, the results are both qualitatively and quantitatively similar to the ones obtained with the degenerate approximation. The largest difference we found is at the surface boundary of the configuration, where due to the low density finite temperature effects are more effective. In Fig.~\ref{fig:neTdeg} we compare the electron density for $r>R_p$ in the degenerate and in the non-degenerate case for $T_\infty = 2.3\times 10^5$ K. For distances $r<R_p$ the results are essentially the same as in the degenerate case. In the region $r<<R_n$ at large densities $> \rho_{\rm nuc} = 2.7\times 10^{14}$ g/cm$^3$, the electrodynamical properties of the configuration i.e. Coulomb potential and electric field remain unperturbed even for very large temperatures $T_\infty \sim 10^{11}$ K. This is due to the fact that thermal effects are largely compensated by the gravitational potential as given by Eq.~(\ref{eq:Tcons}); the Coulomb interaction is not involved in this balance and is not affected by the thermal energy. 

\section{Conclusions}\label{sec:5}

In this article we have addressed three new aspects of the description of a self gravitating system of neutrons, protons and electrons in $\beta$-equilibrium:

1) We have presented the Newtonian limit of the treatment by taking the weak field approximation and the non-relativistic $c \to \infty$ limit of the general relativistic Thomas-Fermi and Einstein-Maxwell equations (\ref{eq:Gab12})--(\ref{eq:GRTF2}). The numerical integration of the Newtonian equations shows that the gravito-electrodynamic structure evidenced in \cite{PLB2011} (see also Sec.~\ref{sec:2}) is not of general relativistic nature but is already present in the Newtonian regime. However, in view of the large quantitative discrepancies of the Newtonian regime in the description of other neutron star properties like mass and radius (see e.g.~Fig.~3 in \cite{merloni98}), a general relativistic treatment is mandatory in any astrophysical consideration.

2) It has been also presented the extension of the previous treatment \cite{PLB2011} to finite temperatures. Although the thermal energy expected to be stored in old neutron stars with surface temperatures $\sim 10^6$ K \cite{crab1,crab2} is much larger than the internal Coulomb energy (see Sec.~\ref{sec:4}), still the electromagnetic structure (see Fig.~\ref{fig:EGRNewtonian}) is unaffected by the presence of the thermal component. Physically this is due to the very large Fermi energy of the neutrons $\sim 1$ GeV, of the protons $\sim 10$ MeV and of the electrons $\sim 0.1$ GeV, as can be seen from Eq.~(\ref{eq:Tfermi}).

3) More generally, we have given the explicit demonstration of the constancy throughout the configuration of the Klein potentials of each species present in the system in the more general case of finite temperatures. This generalizes the condition of the constancy of the general relativistic Fermi energies derived in the special case $T=0$ in \cite{PLB2011}. 

All these results are relevant to the generalization of the Feynman-Metropolis-Teller treatment of compressed atoms to relativistic regimes in presence of thermal effects and they will necessarily also apply in the case of additional particle species and of the inclusion of strong interactions in neutron stars \cite{PLB3011}.


\begin{acknowledgments}
We acknowledge Jim Lattimer for interesting discussions specially on the first two points in the Conclusions at the Les Houches workshop ``From Nuclei to White Dwarfs to Neutron Stars'' (Eds. A.~Mezzacappa and R.~Ruffini, World Scientific 2011, in press). They improved the presentation and made more clear the general validity of our results.
\end{acknowledgments}


\end{document}